\documentclass[twocolumn]{aastex631}

\usepackage{gensymb}

\begin{document}

\title{The first Palomar Gattini-IR catalog of $J$-band light curves: construction and public data release}

\author{Shion Murakawa}
\affiliation{MIT-Kavli Institute for Astrophysics and Space Research, 77 Massachusetts Ave., Cambridge, MA 02139, USA}

\author{Kishalay De}
\altaffiliation{NASA Einstein Fellow}
\affiliation{MIT-Kavli Institute for Astrophysics and Space Research, 77 Massachusetts Ave., Cambridge, MA 02139, USA}

\author{Michael C. B. Ashley}
\affiliation{School of Physics, University of New South Wales, UNSW 2052, Australia}

\author{Nicholas Earley}
\affiliation{Cahill Center for Astrophysics, California Institute of Technology, Pasadena, CA 91125, USA}

\author{Lynne A. Hillenbrand}
\affiliation{Cahill Center for Astrophysics, California Institute of Technology, Pasadena, CA 91125, USA}

\author{Mansi M. Kasliwal}
\affiliation{Cahill Center for Astrophysics, California Institute of Technology, Pasadena, CA 91125, USA}

\author{Ryan M. Lau}
\affiliation{NSF’s NOIRLab, 950 North Cherry Avenue, Tucson, AZ 85719, USA}

\author{Anna M. Moore}
\affil{Research School of Astronomy and Astrophysics, Australian National University, Canberra, ACT 2611, Australia}

\author{J. L. Sokoloski}
\affiliation{Columbia Astrophyiscs Lab, 550 W 120th St, 1027 Pupin Hall, Columbia University, New York, NY 10027, USA}

\author{Roberto Soria}
\affiliation{College of Astronomy and Space Sciences, University of the Chinese Academy of Sciences, Beijing 100049, China}
\affiliation{INAF -- Osservatorio Astrofisico di Torino, Strada Osservatorio 20, I-10025 Pino Torinese, Italy}
\affiliation{Sydney Institute for Astronomy, School of Physics A28, The University of Sydney, Sydney, NSW 2006, Australia}



\begin{abstract}

Palomar Gattini-IR (PGIR) is a wide-field, synoptic infrared time domain survey covering $\approx 15000$\,sq.\,deg. of the accessible sky at $\approx 1-3$\,night cadence to a depth of $J\approx 13.0$ and $\approx 14.9$\,Vega mag in and outside the Galactic plane, respectively. Here, we present the first data release of $J$-band light curves of 2MASS sources within the survey footprint covering approximately the first four years of operations. We describe the construction of the source catalog based on 2MASS point sources, followed by exposure filtering criteria and forced PSF photometry. The catalog contains light curves of  $\approx 286$\,million unique sources with 2MASS magnitudes of $J < 15.5$\,mag, with a total of $\approx 50$\,billion photometric measurements and $\approx 20$\,billion individual source detections at signal-to-noise-ratio $> 3$. We demonstrate the photometric fidelity of the catalog by i) quantifying the magnitude-dependent accuracy and uncertainty of the photometry with respect to 2MASS and ii) comparing against forced PGIR aperture photometry for known variable sources. We present simple filtering criteria for selecting reliable photometric measurements as well as example \texttt{Python} notebooks for users. This catalog is one of the largest compilation of nightly cadence, synoptic infrared light curves to date, comparable to those in the largest optical surveys, providing a stepping stone to upcoming infrared surveys in the coming decade.
\end{abstract}



\section{Introduction} \label{sec:intro}

Catalogs of variability in the optical bands have grown dramatically in the last decade, driven by the development of large optical time domain surveys capable of covering nearly the entire sky every night. Though historically limited and confined to targeted follow-up campaigns, large-scale studies of variability provide avenues to understanding physics ranging from the internal structures of stars \citep{Eyer2008, Catelan2023} to accretion onto super-massive black holes \citep{Ulrich1997, Peterson2001}. In addition, the availability of historical time domain coverage of known sources provides an indispensable baseline for identifying new types of outbursts. Notable examples of catalogs in the optical bands include the All-sky Automated Survey for Supernovae (ASAS-SN; \citealt{Hart2023}), the Asteroid Terrestrial-impact Last Alert System (ATLAS; \citealt{Heinze2018}), the Catalina Real-time Transient Survey (CRTS; \citealt{Drake2009}), PanSTARRS \citep{Chambers2016}, Gaia \citep{Gaia2018}, the Optical Gravitational Microlensing Experiment (OGLE; \citealt{Udalski2003}) and the Zwicky Transient Facility (ZTF; \citealt{Bellm2019}).

The large sky foreground as well as the high cost of detectors have severely limited similar efforts in the infrared bands. Although these efforts have been limited by instrumental sensitivity, infrared variables provide windows to probing the origin of variability in cool and dusty stars, as well as variability in the most dust obscured regions of the universe. While the Two Micron All Sky Survey (2MASS; \citealt{Skrutskie2006}) spearheaded one of the first all-sky static catalogs for near-infrared (NIR) sources, its usage for time domain studies was extremely limited to targeted campaigns (e.g. \citealt{Carpenter2001}). Deeper surveys covering smaller areas of the sky have been carried out in the NIR using the UKIRT and VISTA telescopes. Previous UKIRT surveys covered $\approx 10-20$\,sq.\,deg. of the Galactic bulge searching for microlensing events \citep{Shvartzvald2017}, as well as a up to $\approx 3-4$ epochs of a wider $\approx 500^\circ$ area centered on the northern Galactic plane. Similarly, the VISTA telescope carried out the VISTA Variables in the Via-Lactea (VVV; \citealt{Minniti2010}) survey over $\approx 520$\,sq.\,deg. of the southern Galactic plane and bulge, covering $\sim 10^9$ sources for up to $\approx 100$\,epochs in the $Ks$ band over a duration of $\approx 10$\,years. The resulting dataset was used to produce catalogs of $\approx 490 \times 10^6$ variable star candidates \citep{Molnar_2021, Lopes2020} identified from the VIRAC database \citep{Smith2018} centered on the southern Galactic plane. In the mid-infrared bands, the Spitzer Space Telescope has been used to carry out targeted searches for variables (e.g., \citealt{Freedman2011, Boyer2015, Karambelkar2019}), and the NEOWISE all-sky survey \citep{Mainzer2014} has been used to create six-month cadence light curves of $\approx 1.2 \times 10^9$\,sources, spanning the last $\approx 9$\,years \citep{Meisner2023}.

Palomar Gattini-IR (PGIR) is a ground-based near-infrared time domain survey operating from Palomar Observatory in California \citep{Moore2016, Moore2019} since 2018. Operating in $J$-band with a 25\,sq.\,deg. field of view, PGIR can scan $\approx 15000$\,sq.\,deg. of the accessible night sky with a cadence of $\approx 2$\,nights to a median depth of $J \approx 14.8$\,Vega\,mag\footnote{For the rest of this work, we report all photometry measurements in the Vega magnitude system} ($J \approx 15.7$\,AB\,mag ; \citep{De2020a}) outside the Galactic plane. The combination of depth (comparable to 2MASS) and cadence has already enabled a wide range of time domain science that is difficult with optical surveys due to dust extinction, including the first quantitative constraints on the Galactic nova rate \citep{De2021}, a new census of dusty variables such as R Coronae Variables \citep{Karambelkar2021} and long-period Miras \citep{Suresh2024}, and the discovery of dust enshrouded young stellar outbursts \citep{Hillenbrand2021}. Over the first four years of operations, PGIR has obtained $\gtrsim 100$\,epochs of observations over most of the visible sky, including $\gtrsim 500$ epochs in the northern Galactic plane.

Prior work on variability from PGIR has used forced aperture photometry measurements on targeted samples of sources \citep{De2022} or source detection on stacked images \citep{Suresh2024}, which are limited by the prior target selection criteria and high statistical significance source detection, respectively. In contrast, the 2MASS catalog provides a uniformly selected sample of sources to serve as the baseline catalog for photometry, given the similar depth of the 2MASS and PGIR surveys. In this paper, we present the first public data release of light curves from the PGIR survey obtained by performing forced photometry at the positions of 2MASS point sources over the entire visible sky footprint. Section \ref{sec:selection} describes the source selection criteria, forced photometry technique and computational optimizations. Section \ref{sec:output} describes the output catalog and demonstrates its fidelity in photometric measurements. We provide a summary of the catalog, example use cases and a user-friendly interface in Section \ref{sec:summary}.

\section{Catalog generation}
\label{sec:selection}
\subsection{Input source catalog and data selection criteria}

\begin{figure*}[!ht]
    \centering
    \includegraphics[width=0.95\linewidth]{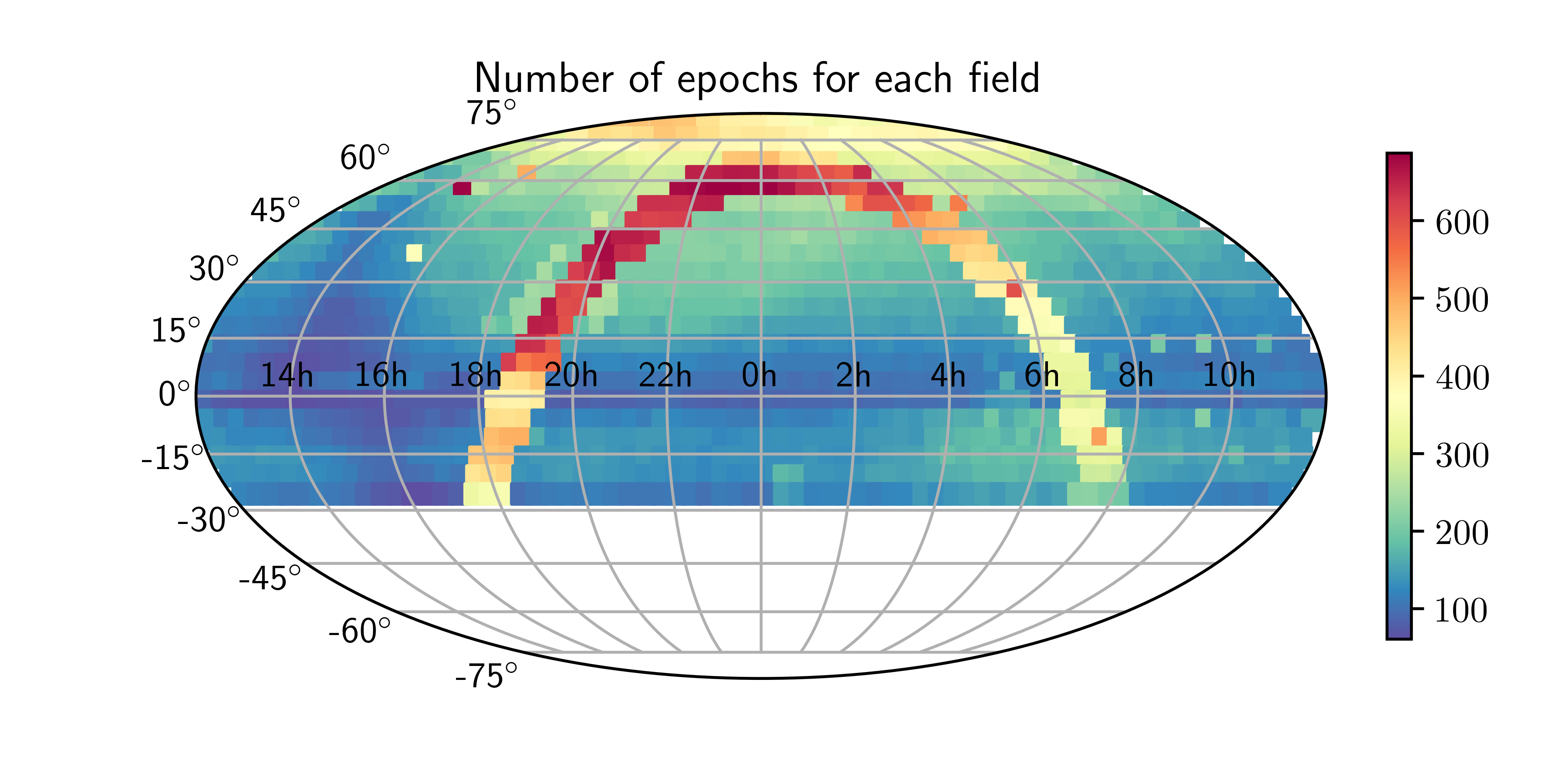}
    \caption{Number of observed epochs per PGIR field in the observing footprint. Fields in the Galactic plane and near the north pole have the most observations due to the adopted survey strategy and visibility from Palomar Observatory.}
    \label{fig:mapplot}
\end{figure*}

The PGIR observing system divides the entire visible sky at $\delta > -29\degree$ into a set of $1329$ fields, with each field spanning a footprint of $\approx 4.96\degree\times4.96\degree$ and an average overlap of $\approx 6$\arcmin\ near the celestial equator (the overlap is larger near the celestial poles). In order to improve data calibration given the known variation of the telescope Point Spread Function (PSF) across the field of view, each field is sub-divided into four quadrants, with each quadrant further divided into four sub-quadrants, resulting in a total of 16 sub-quadrants covering a fixed sky footprint of $\approx 1.24\degree\times1.24\degree$. The corresponding pixel scale is $\approx 4.35$\arcsec\ per pixel, where the images have already been resampled to a $\approx 2\times$ finer pixel scale from the native raw images using the \texttt{Drizzle} algorithm \citep{Fruchter2002}. A full description of the Gattini Data Processing System (GDPS) can be found in \citet{De2020a}. To perform forced photometry, we use the point source catalog from the Two Micron All Sky Survey (2MASS; \citealt{Skrutskie2006}) to select positions of known sources.

Given the typical depth of the PGIR images ($5\sigma$ limiting magnitude of $J \approx 13.0$\,mag and $J \approx 14.8$\,mag in and outside the Galactic plane, respectively; \citealt{De2020a}), we conservatively select all 2MASS sources brighter than $J = 15.5$\,mag within the footprint of a given sub-quadrant. The total sample is $\approx 286 \times 10^6$ sources that satisfy these selection criteria. The time span of the dataset spans the first $\approx 4$\,years of PGIR operations, corresponding to $1475$ nights of formal operations (including down time due to maintenance and weather closures) since first light in October 2018. Whereas the first year of the survey was carried out using a uniform ($\approx 2$\,night) all-sky cadence, the subsequent years have carried out a nightly cadence survey of the visible Galactic plane together with a slower ($\approx 3-4$\,nights) cadence survey for the rest of the sky \citep{De2021}. Figure \ref{fig:mapplot} shows the sky distribution of the number of epochs obtained during this period, demonstrating up to $\approx 700$\,epochs of data in the northern Galactic plane owing to its long visibility from Palomar Observatory. In order to select only photometric quality images, we require that the scatter in the zero-point measurement of the stacked image is $< 0.25$, based on empirical estimates of the photometric quality in poor observing conditions (e.g., during high humidity or through clouds).

\subsection{Forced photometry}

We perform forced Point Spread Function (PSF) photometry at the positions of the 2MASS sources selected in each field. We chose PSF photometry as the optimal option to account for the large variations in the PSF as a function of position in the focal plane \citep{De2020a}. We use the PSF models generated for each sub-quadrant as part of the GDPS, represented by a $15 \times 15$\,pixel model from the resampled image. We multiply the normalized (i.e., summed to unity) PSF model with the science image at the position of each source to calculate the PSF-matched optimal flux of the source as 
\begin{equation}
 F =\frac{\sum_i P_i S_i}{\sum P_i^2},
\end{equation}
where $P_i$ is the PSF model, $S_i$ is the science image and $F$ is the source flux. We converted the flux to physical units using the image zero-point stored in the image headers (and provided in the output catalog). Uncertainties on this flux are computed using two techniques. First, we compute the nominal statistical uncertainty on the flux using
\begin{equation}
    \sigma_F = \frac{\sqrt{\sum_i P_i^2 \sigma_i^2}}{\sum{P_i^2}},
\end{equation}

where $\sigma_i^2$ is the pixel variance derived by summing the variance in the pixel values (assuming Poisson statistics) across the PSF footprint, and $\sigma_F$ is the statistical flux uncertainty. As the uncertainty can be underestimated from statistical noise alone (e.g., due to the presence of confusion noise, or correlated pixel noise), we also estimate the flux uncertainty through a Monte-Carlo procedure by computing the flux $F$ at 100 random positions in the image and providing the standard deviation of these fluxes in the output catalog.

\subsection{Parallelization}

We parallelize the forced photometry computation by using a total of 28 (out of 32) cores on the GDPS compute node (see Figure 2 in \citealt{De2020a}). Given the large number of sources in each sub-quadrant due to the field size (especially near the Galactic plane), the creation of the catalog is limited by the system memory rather than compute power. We therefore simultaneously execute the catalog creation for multiple sub-quadrants until the system memory limit is reached (the amount of memory required per sub-quadrant depends on the number of sources, and hence field sky position). A new set of sub-quadrants are executed only after the complete set in a given execution call are completed. As the GDPS compute node is also responsible for nightly data time processing, the catalog creation is paused every night once the nightly data are received from the observatory, and automatically re-initiated after the end of nightly operations.

\section{Output catalog}
\label{sec:output}
\subsection{Storage schema and quality flags}
\label{sec:schema}

\begin{table*}[!ht]
    \centering
    \begin{tabular}{|c|c|c|}
    \hline
    \hline
    Column name & Data Type & Description\\
    \hline
    {\bf Exposures} & &\\
    stackquadid & \texttt{int32} & Unique exposure index \\
    obsjd & \texttt{float32} & Julian date of exposure start\\
    exptime & \texttt{float32} & Total exposure time in seconds \\
    airmass & \texttt{float32} & Exposure Airmass \\
    zp & \texttt{float32} & Exposure zero point magnitude \\
    zp\_rms & \texttt{float32} & Exposure zero point root mean sqaure devitation \\
    zp\_unc & \texttt{float32} & Exposure zero point uncertainty \\
    limmag & \texttt{float32} & Average $5\sigma$ limiting magnitude \\
    color\_term & \texttt{float32} & Color term in 2MASS $J-H$ color \\
    color\_unc & \texttt{float32} & Uncertainty in $J-H$ color term\\
    saturmag & \texttt{float32} & Nominal exposure saturation magnitude \\
    numnoisepixels & \texttt{float32} & Effective footprint of PSF in number of pixels \\
    mside & \texttt{object} & East or West, side of pier meridian of exposure \\
    nightid & \texttt{int32} & Indexed night number of exposure \\
    filename & \texttt{object} & File name of original image \\
    
    \hline 
    {\bf Sources} & & \\
    tmcra & \texttt{float32} & 2MASS right ascension of source \\
    tmcdec & \texttt{float32} & 2MASS declination of source \\
    pts\_key & \texttt{int64} & 2MASS ID of source \\
    tmcjmag & \texttt{float32} & 2MASS $J$-band magnitude \\
    tmchmag & \texttt{float32} & 2MASS $H$-band magnitude \\
    psfcontam & \texttt{int32} & Number of other sources within the source's PGIR PSF aperture \\
    meanmag & \texttt{float32} & Mean magnitude of all PGIR exposures of source \\
    maxmag & \texttt{float32} & Maximum magnitude of all PGIR exposures of source \\
    minmag & \texttt{float32} & Minimum magnitude of all PGIR exposures of source \\
    rmsmag & \texttt{float32} & Magnitude RMS of all PGIR exposures of source \\
    
    \hline
    {\bf Sourcedata} & & \\
    stackquadid & \texttt{int32} & Exposure identification \\
    obsjd & \texttt{float32} & Julian date of exposure start \\
    pts\_key & \texttt{int64} & 2MASS ID number of source \\
    magpsf & \texttt{float32} & $J$-band magnitude of PGIR exposure of source \\
    magpsferr & \texttt{float32} & Image noise magnitude error of PGIR exposure of source \\
    magpsfstaterr & \texttt{float32} & Statistical magnitude error of PGIR exposure of source \\
    magpsflim & \texttt{float32} & Estimated $3\sigma$ limiting magnitude at source position \\
    flags & \texttt{int32} & Bit-value of the flags of exposure of source \\
    \hline
    \end{tabular}
    \caption{Column names and descriptions of the HDF5 matchfile for each sub-quadrant (see text).}
    \label{tab:storage table}
\end{table*}

\begin{figure*}[!ht]
    \centering
    \includegraphics[width=0.7\linewidth]{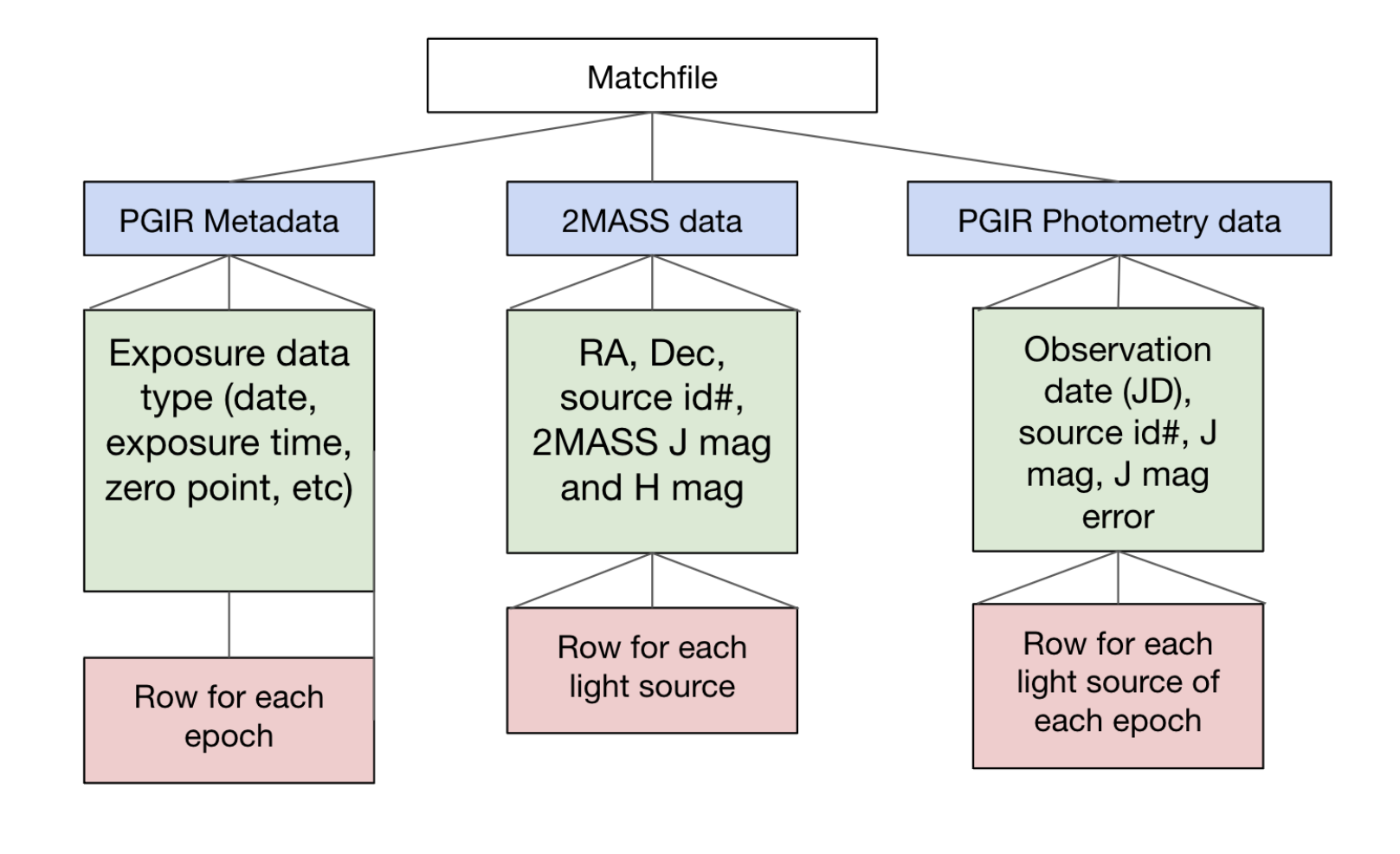}
    \caption{Schematic depiction of the contents of a matchfile.  Blue boxes indicate the sub-tables, green boxes indicate the data columns (column descriptions are provided in Table \ref{tab:storage table}), and red boxes clarify the distinction between each row in the tables.}
    \label{fig:storage schema}
\end{figure*}

The photometry catalog and metadata for each field sub-quadrant are stored in a single file, henceforth referred to as a `matchfile' as it is created by performing photometry on matched sources across all the epochs of observation. The complete catalog consists of a total of 21264 matchfiles spread over the observing footprint, stored in the HDF5 format. Table \ref{tab:storage table} shows the details of the contents of each matchfile, and Figure \ref{fig:storage schema} is a schematic that displays the data structure of the matchfiles, which may be helpful when accessing the photometric data and associated exposure metadata. Each matchfile consists of three subtables:
\begin{itemize}
    \item \texttt{Exposures}: Metadata for each good quality exposure that contributed to the photometry contained in the matchfile.
    \item \texttt{Sources}: A summary of the 2MASS photometric properties for each point source that was included in the photometry catalog, in addition to statistics quantifying the photometric variability of the source measured by PGIR in the catalog.
    \item \texttt{Sourcedata}: Individual PGIR photometric measurements for each point source in the catalog, in addition to quality flags for each measurement. 
\end{itemize}

In addition to the nominal photometric measurements derived from the images, we provide a number of quality flags in the \texttt{Sourcedata} table to allow users to filter on specific measurements that may be deemed to be unreliable. The quality flag (\texttt{flags} in Table \ref{tab:storage table}) is stored as a single base-10 representation of a 6-bit binary flag, consisting of the following boolean qualifiers ordered from lowest to highest bit:

\begin{enumerate}
 \item \texttt{F1}: If the exposure was acquired on the west side of the telescope meridian axis. Because of the equatorial mount, the same source may be observed on different sides of the mount, changing the PSF and causing systematic offsets in photometric measurements. As most observations are acquired on the `East' side of the mount, users may choose to only select observations from one mount side to avoid systematic offsets. 
 \item \texttt{F2}: If the source magnitude is brighter than the nominal exposure saturation magnitude. The nominal saturation magnitude of the PGIR system was $J \approx 8.5$\,mag between 2018 and 2020. Due to an improvement in the readout system, the saturation magnitude was improved to $J\approx 6.0$\,mag subsequently. Users to may choose to avoid measurements for variable sources that are brighter than the nominal saturation threshold for a particular epoch.
 \item \texttt{F3}: If the source magnitude is outside of the recommended reliable magnitude range for the epoch. Due to heavy source confusion near the Galactic plane, we identify reliable magnitude ranges for photometry across the sky. Given the pixel scale in the drizzled images, there will be on average more than two sources within a PSF footprint if the number of sources within a sub-quadrant exceeds $10^4$. Therefore, we identify sub-quadrants with $> 10^4$ sources as Galactic plane fields, and recommend that photometry is only reliable (better than 10\%) for sources between the saturation magnitude and $J < 13$\,mag at a given epoch ($\gtrsim 10\times$ brighter than the nominal source selection threshold). For other fields, the photometry may be reliable for sources up to the $J = 15.5$\,mag threshold.
 \item \texttt{F4}: If the airmass of the exposure from which the photometry was obtained is $>2$. More than $90$\% of PGIR exposures are obtained at airmass $<2$ \citep{De2020a}, beyond which the image quality degrades. We therefore flag all photometry obtained from higher airmass exposures.
 \item \texttt{F5}: If the photometric zero-point of the exposure deviates by more than $0.75$\,mag from the median zero-point of all exposures for that sub-quadrant. Low quality exposures obtained in poor weather conditions typically exhibit zero-point magnitudes deviant from the average zero-point for the field. We flag photometry measurements obtained from exposures where the zero-points are significantly different (corresponding to $\approx 2\times$ in flux transparency) from the median zero-point.
 \item \texttt{F6}: If there are any additional sources in the PSF footprint of a source that have 2MASS magnitudes brighter than $M + 2$\,mag, where $M$ is the measured PGIR magnitude of the source for a given epoch. As source confusion is expected to be a limitation, but not as constraining for bright variable sources, we include this flag to tag photometry measurements where there are other sources within the PSF footprint brighter than $M + 2$\,mag. For example, if the source is nominally faint in the Galactic plane ($J \gtrsim 13$\,mag, fainter than the detection threshold of \texttt{F3}) but there are no other sources in the PSF aperture to $<15.5$\,mag, users may choose to use the source photometry. 
\end{enumerate}

\subsection{Photometric accuracy}

\begin{figure*}
    \includegraphics[width=0.49\textwidth]{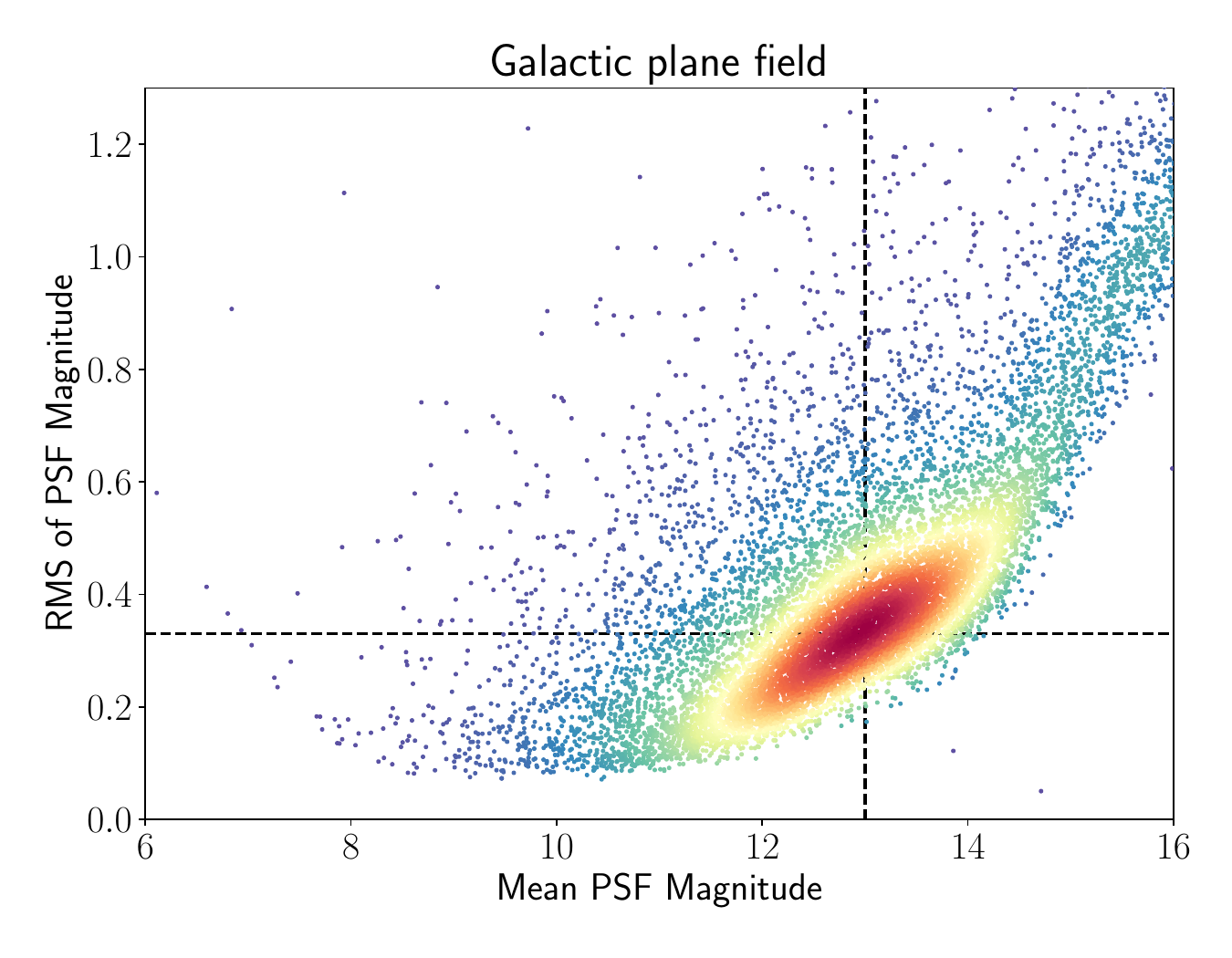}
    \includegraphics[width=0.49\textwidth]{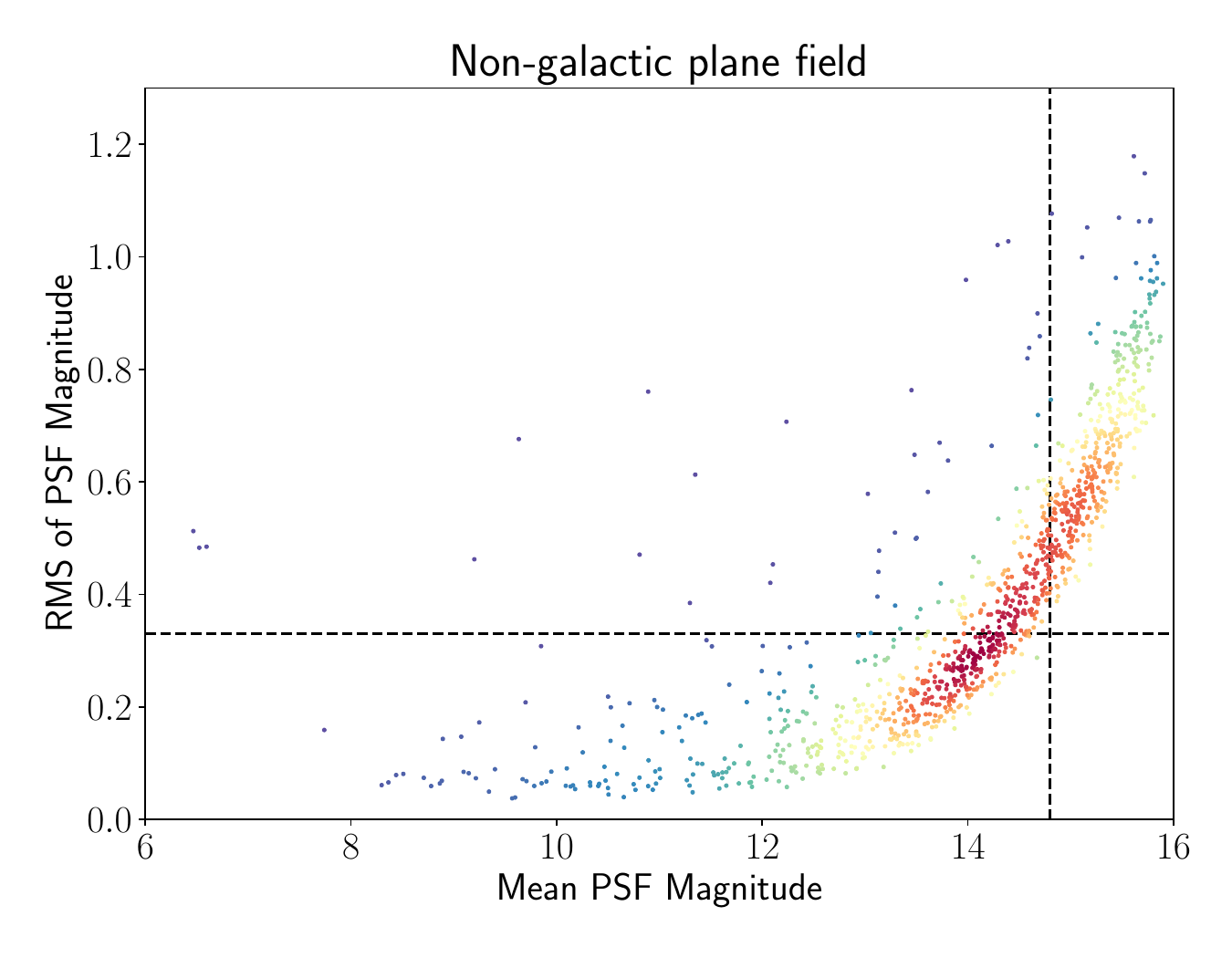}
    \caption{Hess diagram showing the magnitude-dependent scatter of PGIR photometry across all epochs for a representative Galactic plane (left) and extragalactic (right) field sub-quadrant. Each circle represents a unique source in the field, which redder regions representing areas of higher source density. The horizontal dashed line marks a photometric scatter of $\approx 0.33$\,mag, corresponding to a SNR of $\approx 3$. The vertical dashed line shows the estimated faintest reliable magnitude suggested for a typical Galactic ($\approx 13.0$\,mag) and extragalactic ($\approx 14.8$\,mag) field. Note that some sources have mean magnitudes of $J > 15.5$\,mag (despite the sample selection) due to fainter fluxes over the PGIR observing period than 2MASS.}
    \label{fig:rms_plot}
\end{figure*}

The photometric accuracy of the PGIR observing system has been previously published in \citet{De2020a}, demonstrating the photometric stability in both dense and sparse fields. Here, we demonstrate the photometric fidelity of the PGIR matchfile photometry by direct measurements of the photometric scatter as well as comparing against the 2MASS catalog, both with and without the photometric flags. In Figure \ref{fig:rms_plot}, we first show the magnitude-dependent scatter of photometric measurements for a typical Galactic and extragalactic field. The photometric scatter systematically increases to fainter magnitudes as expected, and it notably higher in the Galactic plane field at a fixed magnitude due to source confusion. Figure \ref{fig:rms_plot} also shows the suggested faintest magnitude for reliable measurements in both types of fields, where the photometric scatter corroborates the suggested range. In addition, Figure \ref{fig:rms_plot} shows that the best photometric precision achieved is $\lesssim 5$\% for sources brighter than $\approx 12$\,mag. While the faintest sources included in the catalog are nominally fainter than the suggested range, they are included for completeness in case they are of interest during brief flaring episodes.

\begin{figure}[!ht]
    \centering
    \includegraphics[width=1\linewidth]{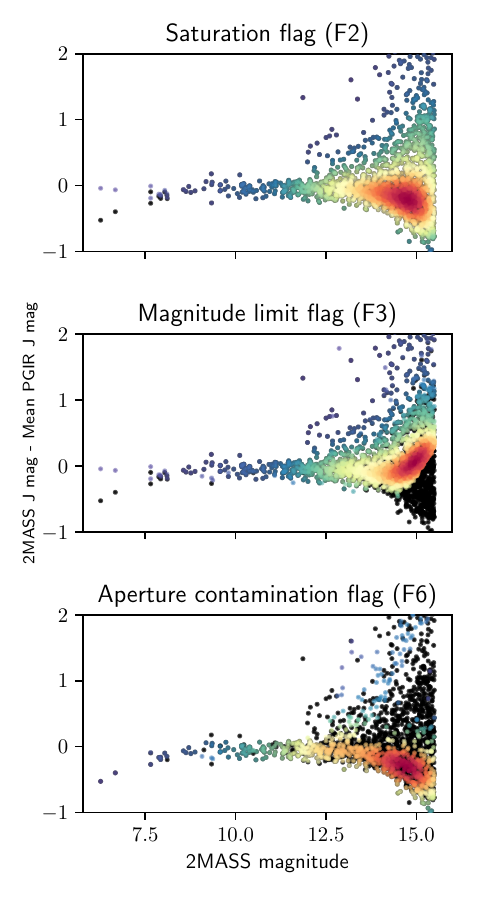}
    \caption{Application of flags to PGIR photometry to exclude potentially unreliable measurements in a crowded Galactic plane field. In each panel, black points show the difference between reported 2MASS $J$-band photometry and the mean PGIR magnitude measured across all epochs included in the match file, without any filtering. Plotted on top of the black points, the color shaded points show the same quantity after excluding measurements that are flagged to be unreliable (i.e., the flag is set to True) based on the condition shown in the panel title, with regions of redder color indicating higher source density.}
    \label{fig:flags}
\end{figure}

Both scatter plots show a substantial population of high scatter sources with variability significantly exceeding the global trend with magnitude, confirming the rich landscape of variables in the catalog. We also observe an increase in scatter for sources brighter than $\approx 8.5$\,mag due to saturation of sources in the earlier epochs of the survey; these measurements can be excluded with the use of the catalog flags. To demonstrate this utility, we show in Figure \ref{fig:flags}, the difference between the mean PGIR magnitude measured from the matchfile photometry (including all measurements) and the published 2MASS magnitudes as a function of the source brightness for a Galactic plane field. Consistent with Figure \ref{fig:rms_plot}, the global scatter with respect to 2MASS (shown with the black points) increases towards fainter magnitudes. We show the effect of the application of flags to exclude unreliable measurements affected by saturation (\texttt{F2}) and confusion (\texttt{F3} and \texttt{F6}) with colored points in Figure \ref{fig:flags}.

The saturation flag (\texttt{F2}) produces improved photometry for bright sources ($J < 8.0$\,mag) by excluding epochs (in the first year of the survey) in which sources at $8.5 < J < 6.0$ would be saturated. The magnitude limit flag (\texttt{F3}) also excludes measurements for saturated sources on the bright end, while also excluding measurements that are fainter than $J = 13$\, mag for galactic plane fields (and $J = 15.5$\,mag for non-galactic plane fields) for faint sources near the detection threshold, thereby biasing the mean measured magnitude to be higher than the 2MASS photometry and producing the upward trend. The aperture contamination flag (\texttt{F6}) produces a similar trend -- sources that have another contaminating source within the PSF aperture would typically appear brighter than 2MASS when including all measurements, but when excluding measurements during which the contamination is significant, only measurements when the source $> 2$\,mag brighter than the contaminants are included in the mean measurement, producing the positive tail of the residuals. 

\begin{figure*}[!ht]
    \centering
    \includegraphics[width=0.49\textwidth]{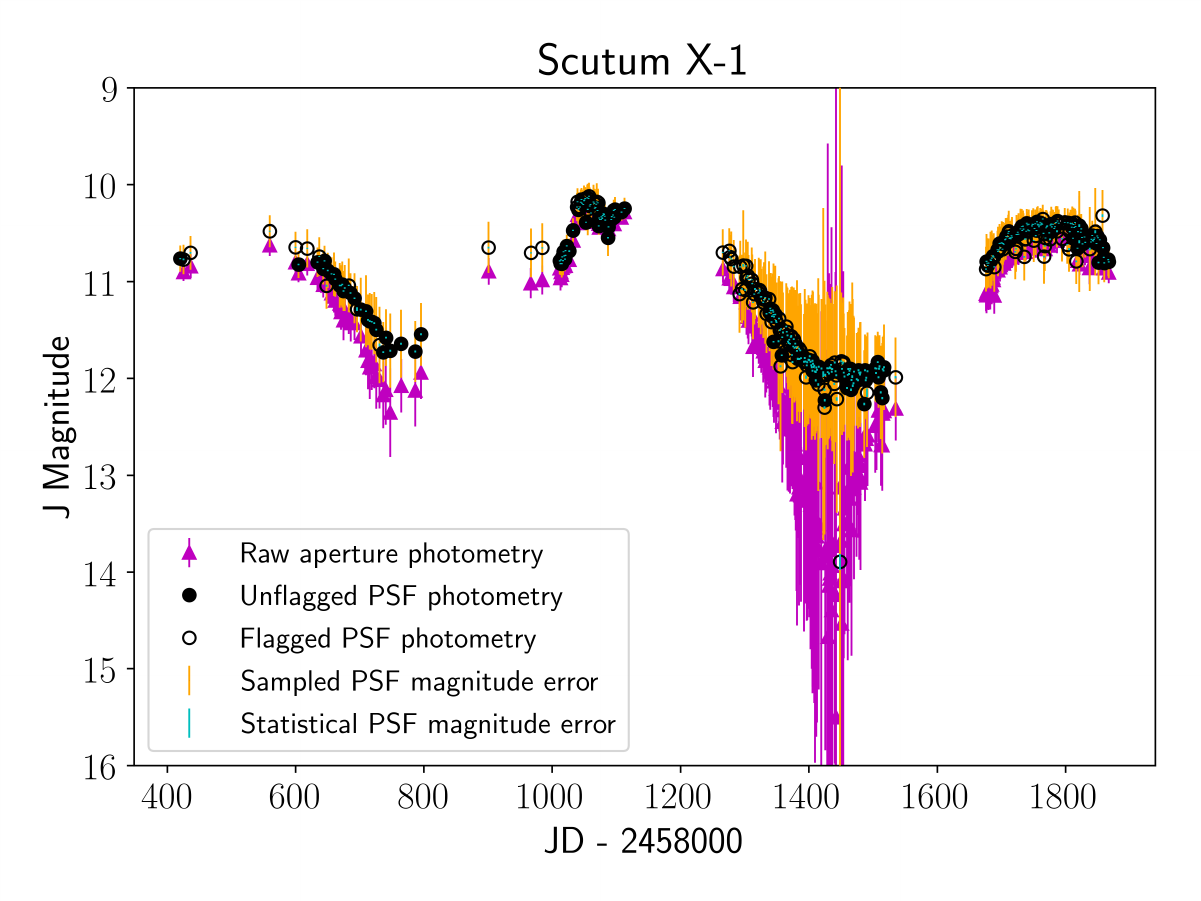}
    \includegraphics[width=0.49\textwidth]{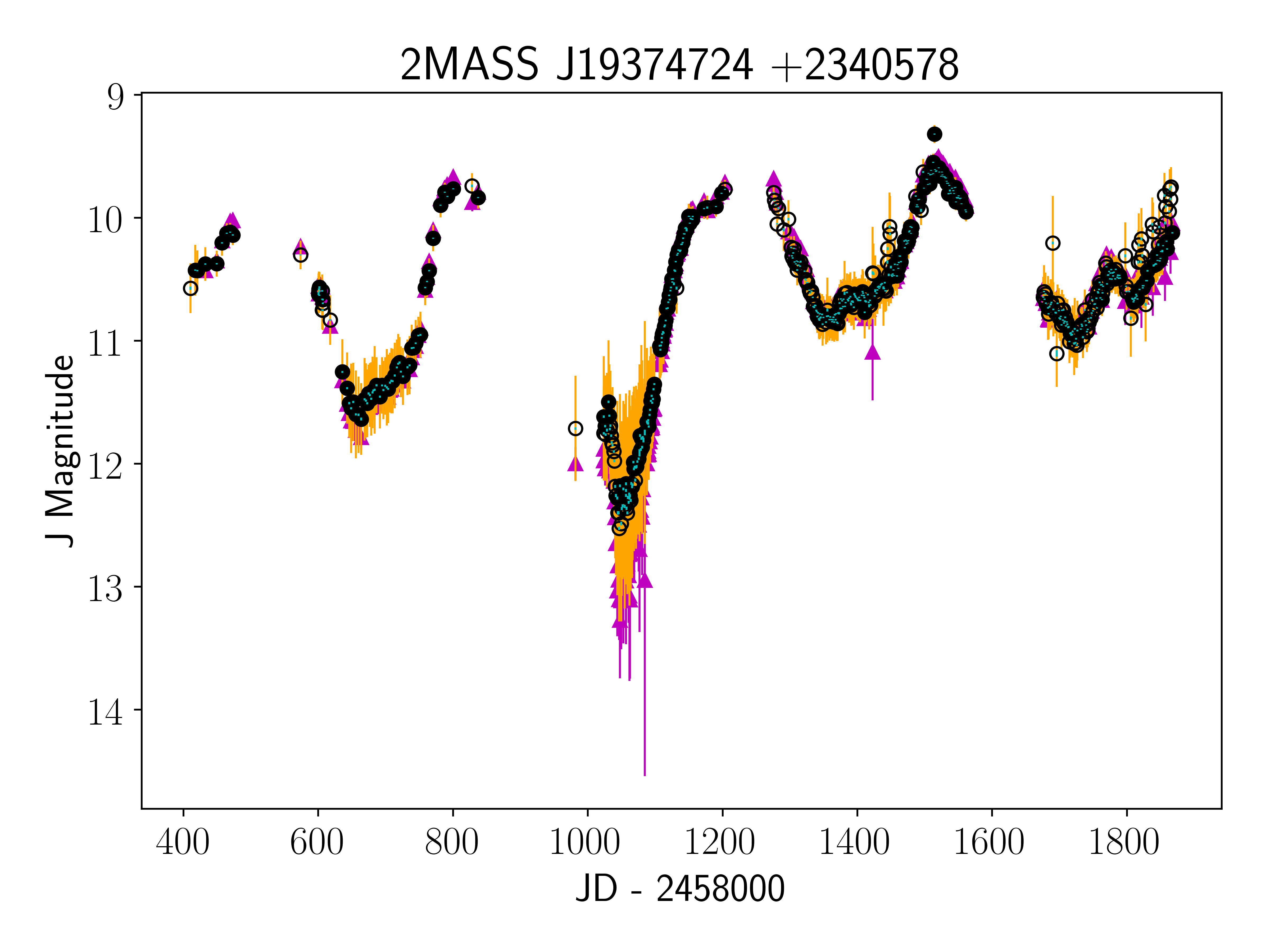}
    \includegraphics[width=0.49\textwidth]{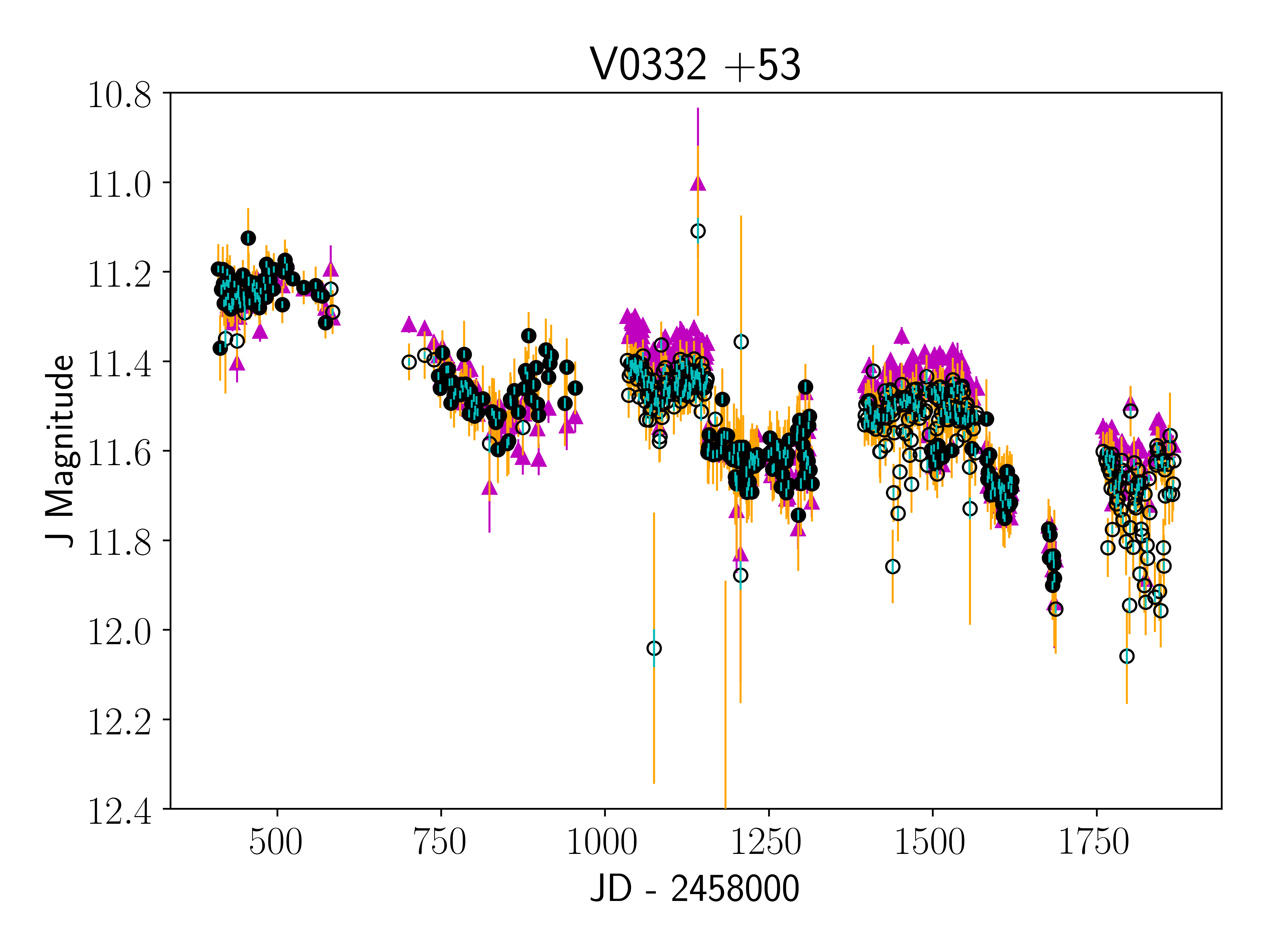}
    \includegraphics[width=0.49\textwidth]{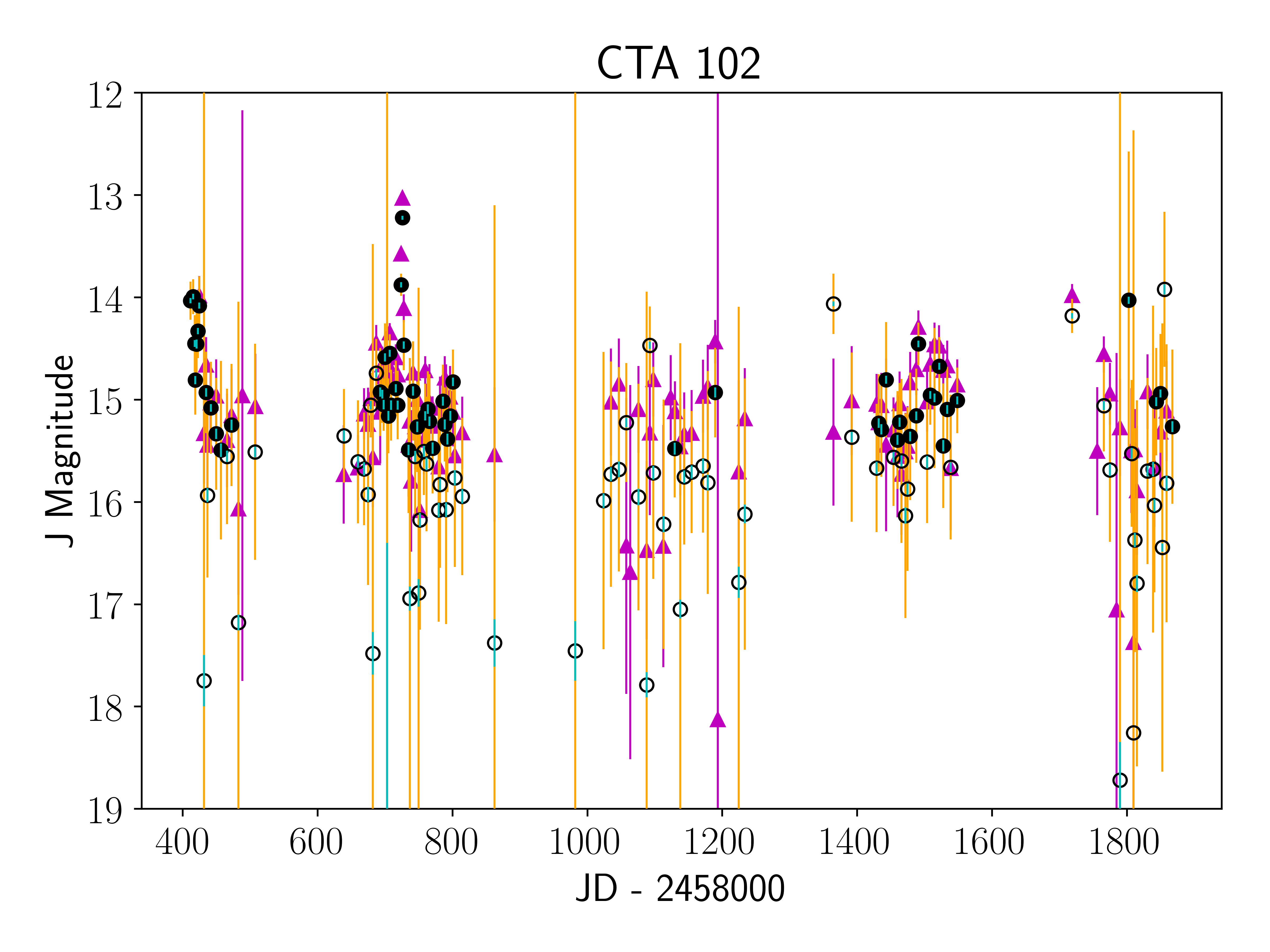}
    \includegraphics[width=0.49\textwidth]{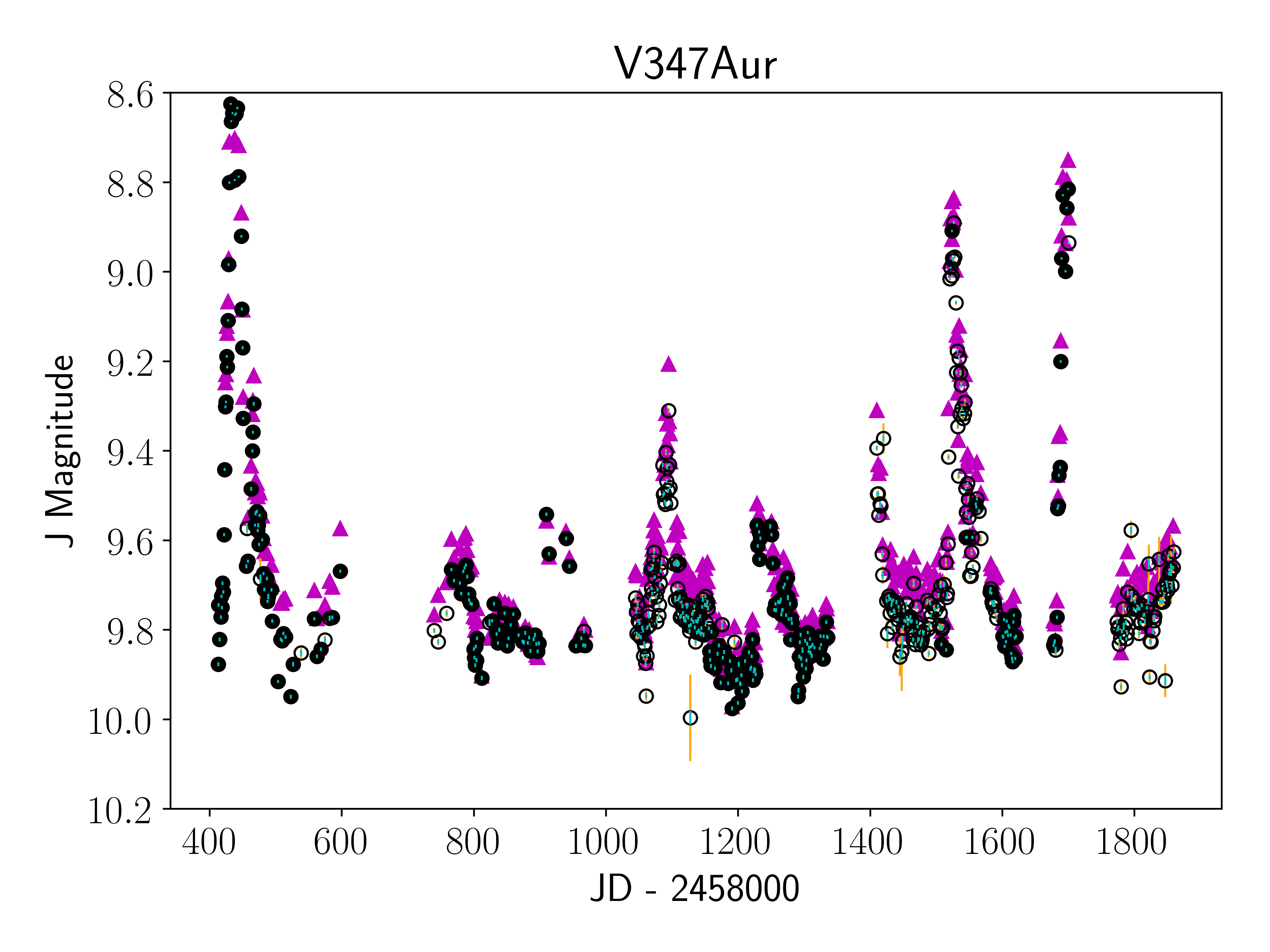}
    \includegraphics[width=0.49\textwidth]{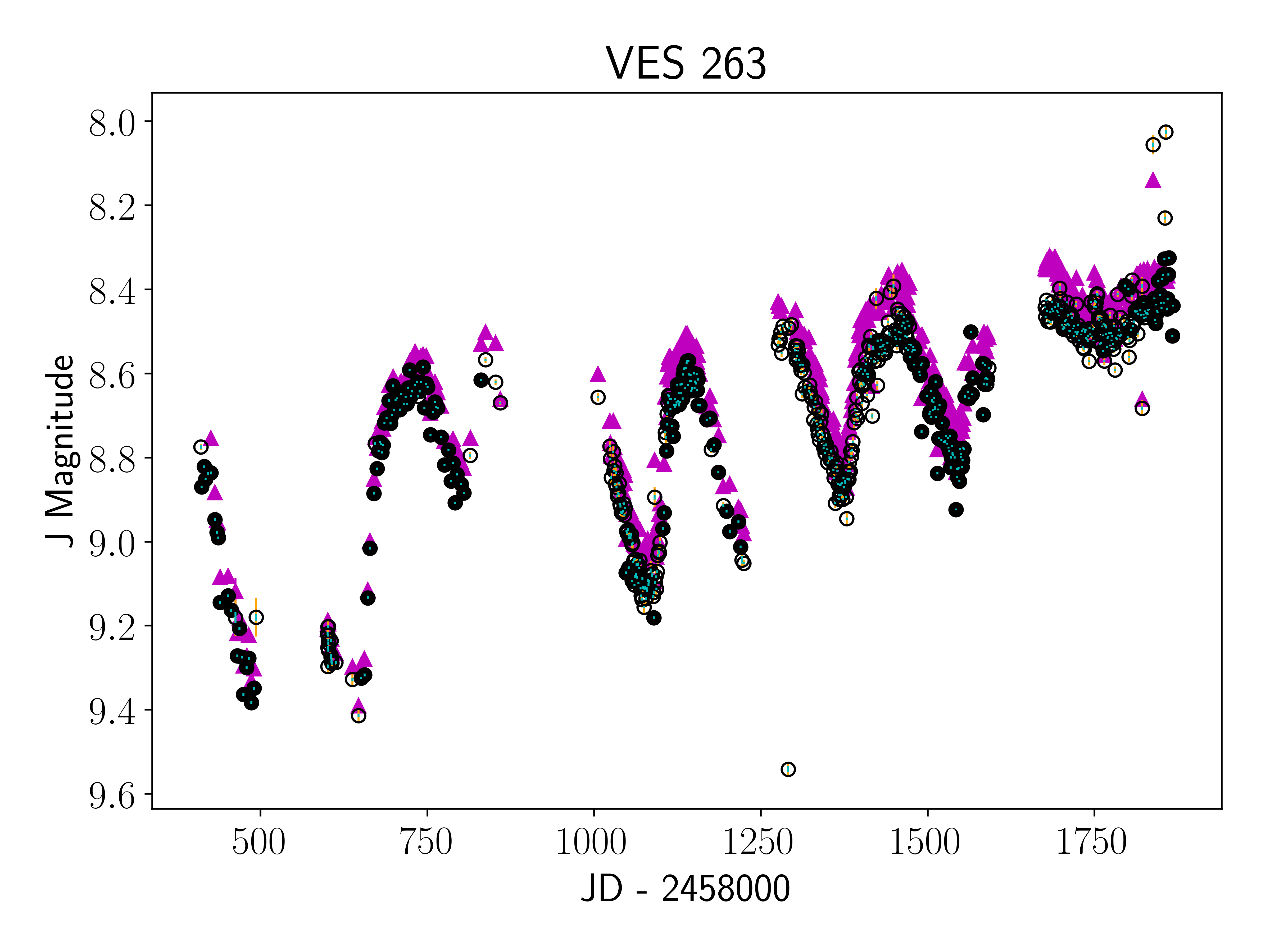}
    \caption{Example light curves derived from the PGIR PSF forced photometry catalog (in black) with statistical errors (in cyan) and errors estimated from sampling across the image (in orange). PSF photometry are divided into ones deemed to be reliable (unflagged, filled circle) and potentially erroneous (flagged, hollow circle). For comparison, we also show forced aperture photometry (in pink) at the source location. {\it Top Left}: The periodic pulsations observed in the heavily obscured Mira variable X-ray binary, previously studied with aperture photometry in \citet{De2022}. {\it Top Right}: Same as the left panel for multi-modal pulsations observed in a long-period variable identified in the catalog of \citet{Suresh2024}. {\it Middle Left}: Long term decay superimposed with shorter period pulsations in the Be X-ray binary V0332+53. {\it Middle Right}: Sporadic flaring activity in the known blazar CTA\,102. {\it Bottom Left}: The young star candidate V347\,Aur exhibiting quasi-periodic outbursts \citep{Dahm2020}. {\it Bottom Right}: The classical Be star VES\,263 undergoing periodic outbursts \citep{Froebrich2023}.} 
    \label{fig:examples}
\end{figure*}

Next, we compare light curves produced by the legacy aperture photometry pipeline in PGIR (described in \citealt{De2020a}) to the PSF photometry in this catalog, and further demonstrate the utility of the flags in filtering potentially erroneous photometry. In Figure \ref{fig:examples}, we show historical light curves of some known infrared variable sources. The light curves generally show excellent agreement between the PSF and aperture photometry methods. We note that the PSF photometry measurements show larger fluxes as the sources become faint; we attribute this effect to the explicit annular background subtraction that is carried out in the aperture photometry pipeline to remove nearby contamination. As this type of subtraction is not universally reliable in complicated backgrounds\footnote{\url{https://wise2.ipac.caltech.edu/staff/fmasci/ApPhotUncert.pdf}}, we do not include it in the PSF photometry method. In addition, the PSF photometry method is better suited for photometric measurements as it incorporates the complex PSF shape variation across the focal plane, unlike the aperture photometry method, in which the size of the circular aperture has to be manually chosen to match the PSF footprint. 

The PSF photometry catalog includes uncertainty estimates derived from both the expected statistical (Poisson) noise as well as that estimated from sampling photometry across the image (which is better suited to account for confusion in dense fields). We show both the types of estimates in Figure \ref{fig:examples}. The confusion noise is substantial for sources in dense Galactic plane fields and likely overestimates the true magnitude error for bright sources (e.g. in the case of Scutum X-1 and 2MASS\,J19374724+2340578). For sparse fields, the statistical noise is generally in agreement with the sampled noise estimate for both bright (e.g. V0332+53) and faint (e.g. CTA\,102) sources. The availability of both estimates provides users the flexibility to choose either option based on the application. In addition, Figure \ref{fig:examples} also shows photometry are flagged as being potentially suspect based on our filtering criteria, and effectively eliminates the worst outliers that affect both PSF and aperture photometry (e.g. in V0332+53 and V347\,Aur). We also note the limiting case of CTA\,102 which is sporadically variable and has been mostly remained fainter than the nominal $J = 15.5$\,mag threshold during PGIR operations. Figure \ref{fig:examples} shows that application of all the flags (in particular, \texttt{F3}) for such sources removes photometry measurements when the source is faint (producing gaps in the light curve) and should be used accordingly.

\section{Summary}
\label{sec:summary}
In this paper, we present the construction and schematics of the first catalog of $J$-band light curves from the PGIR survey. The catalog includes $J$-band photometry for $\approx 286 \times 10^6$ 2MASS point sources brighter than $J = 15.5$\,mag in the PGIR observing footprint. Compiled over the entire footptint, we estimate a total of $\approx 50 \times 10^9$ photometric measurements in the catalog, including $\approx 20 \times 10^9$ high significance detection of sources with signal-to-noise ratio $>3$. As such, this catalog represents the first systematic catalog of NIR light curves from a synoptic survey covering $\approx 3/4$ of the sky at few-night cadence, comparable to the largest optical time domain surveys. For comparison, the ATLAS catalog of optical light curves consists of $\approx 30\times 10^9$ measurements \citep{Heinze2018}, while the latest release of the ZTF light curve catalog (\citealt{Masci2019}; DR21) contains $\approx 800 \times 10^9$ measurements\footnote{\url{https://irsa.ipac.caltech.edu/data/ZTF/docs/releases/ztf_release_notes_latest}}. Unlike prior work on PGIR data using targeted forced photometry measurements, this pre-defined and complete catalog offers the possibility of directly searching for specific types of variables using photometric selection techniques. We demonstrate the photometric quality of the catalog and utilities of the included flags via examples from the output catalog as well as light curves of known regular and sporadically variable sources.

A user-friendly \texttt{Python} notebook designed to extract the light curve of a single source is available online\footnote{\url{https://github.com/dekishalay/pgirdps/blob/master/matchfiles/PGIR_make_lightcurve.ipynb}}, and may serve as a useful example for working with the light curve data. The entire catalog of match files (one \texttt{HDF5} file for each sub-quadrant in each field, totalling 21264 files) will be made available as a tarball online. The catalog will also be made available via a query-able interface on NOIRLab's Astro Data Lab Service\footnote{\url{https://datalab.noirlab.edu/PGIR}}. PGIR continues to operate into the ongoing era of deeper NIR time domain surveys such as WINTER \citep{Lourie2020} and PRIME \citep{Kondo2023}, and planned to overlap with the {\it Rubin} observatory \citep{Ivezic2019} that will provide exquisite multi-color optical coverage (to $\lesssim 1\,\mu$m). Therefore, the long temporal baseline of measurements included in this catalog and its subsequent releases will provide a stepping stone to rich time domain investigations in the dynamic infrared sky, in preparation for the exceptionally deep and high cadence infrared surveys of the {\it Roman} space telescope \citep{paladini2023romanearlydefinitionastrophysicssurvey}.

\section*{Acknowledgments}
Palomar Gattini-IR (PGIR) is generously funded by Caltech, Australian National University, the Mt Cuba Foundation, the Heising Simons Foundation, the Binational Science Foundation. PGIR is a collaborative project among Caltech, Australian National University, University of New South Wales, Columbia University and the Weizmann Institute of Science. K. D. was supported by NASA through the NASA Hubble Fellowship grant \#HST-HF2-51477.001 awarded by the Space Telescope Science Institute, which is operated by the Association of Universities for Research in Astronomy, Inc., for NASA, under contract NAS5-26555. MMK acknowledges generous support from the David and Lucille Packard Foundation. JLS acknowledges NSF grant AST-1816100. MMK acknowledges the US-Israel Bi-national Science Foundation Grant 2016227. MMK and JLS acknowledge the Heising-Simons foundation for support via Scialog fellowships of the Research Corporation. MMK and AMM acknowledge the Mt Cuba foundation.


\vspace{5mm}

\bibliography{sample631}{}
\bibliographystyle{aasjournal}



\end{document}